\documentstyle[12pt,aaspp4,epsfig]{article}
\def\paren#1{\left( #1 \right)}

\begin{document}
\title{Light Curves of GRB Optical Flashes}
\author{Shiho Kobayashi}
\affil{Department of Earth and Space Science, Osaka University, Toyonaka
560, Japan} 

\begin{abstract}
The standard model of GRB afterglows assumes that relativistically
expanding material is decelerating due to interaction with the
surrounding medium. The afterglows are well described by the
synchrotron radiation from a forward shock, while the strong optical
flash associated with GRB~990123 can be attributed to the emission 
from a reverse shock. We give a detailed study on the reverse shock
emission. The full light curves are calculated for a long and a short
 GRB cases. We discuss the lack of the prompt optical detections by
 ROTSE for GRB~981121 and GRB~981223.       
\end{abstract}
\keywords{gamma rays: bursts; hydrodynamics; shock waves; relativity}
\section{Introduction}
The gamma-ray burst (GRB) afterglow is believed to involve a
relativistically expanding fireball. The surrounding matter, which we
will refer to as the ISM, influences the fireball shell after it has
been collected enough and the considerable energy has been transferred
from the shell to the ISM. The energy transfer is due to two shocks: a
forward shock propagating into the ISM and a reverse shock propagating
into the shell.  

The afterglow observations are fairly well described by the synchrotron
emission from the ISM electrons accelerated by the forward shock, it is
considered as a confirmation of the relativistic fireball. However, the
current afterglow observations detect the radiation from several hours
after the GRBs. At this stage, the Lorentz factor of the forward shock
is not ultra-relativistic, less than $\sim 10$. Furthermore, the
dynamics depends only on two parameters: the explosion energy and the
ISM density. The afterglow observations provide neither verification of
the extreme relativistic motion nor the properties of the fireball which
can constrain models of the GRB source. 

The counterpart of the afterglow, the emission from a reverse shock
was also predicted (M\'{e}szaros \& Rees 1997; Sari \& Piran 1999a).
When a reverse shock crosses a shell, the forward shocked ISM 
and the reverse shocked shell carry comparable amount of
energy. However, the typical temperature of the shocked shell is
lower since the mass density of the shell is higher. Consequently, the
typical frequency from the shocked shell is lower. A prompt
optical emission from GRB~990123 (Akerlof et al. 1999) can be regarded
as this emission (Sari \& Piran 1999b; Kobayashi \& Sari 2000). 

The emission from the reverse shock is sensitive to the initial
properties of the fireball. The observations can provide some important
clues on the nature of the GRB source. Previous studies focused on the
emission at the peak time. In this paper we calculate the 
full light curves for several frequency regimes. We in section 2
discuss the hydrodynamics of the relativistic fireballs on which the
light curves highly depend. In section 3 we calculate the light curves
for a long burst and a short burst case. We compare our estimates with
the ROTSE observations in section 4. We estimate the initial
parameters of the fireball of GRB~990123, and then we make some comments
on the lack of the prompt detections by ROTSE for GRB~981121 and
GRB~981223. In section 5 we give conclusions.    

\section{Hydrodynamics of a Relativistic Shell}

Consider a relativistic shell with an energy $E$, a Lorentz factor
$\eta$ and a width in laboratory frame $\Delta_0$ expanding into a
surrounding medium (ISM) with a particle number density $n_1$. When the
shell sweeps a large volume of the ISM, it begins to be decelerated. The
interaction between the shell and the ISM is described by two shocks: a
forward shock propagating into the ISM and a reverse shock propagating
into the shell. There are four region separated by the two shocks: the
ISM(denoted by the subscript 1), the shocked ISM(2), the shocked shell
material (3) and the unshocked shell material (4). Using the jump
conditions for the shocks and the equality of pressure and velocity
along the contact discontinuity, we can estimate the Lorentz factor
$\gamma$, the pressure $p$ and the number density $n$ in the shocked
regions as functions of three variables $n_1, n_4$ and $\eta$
(Blandford and McKee 1976). 

There are two limits to get a simple analytic solution (Sari and Piran
1995). If the shell density is high $n_4 \gg \eta^2 n_1$, the reverse
shock is Newtonian which means that the Lorentz factor of the shocked 
shell material $\bar{\gamma}_3$ is almost unity in frame of the
unshocked shell material. It is too weak to slow down the 
shell effectively $\gamma_3 \sim \eta$. On the other hand, if the
density is low $n_4 \ll \eta^2 n_1$, the reverse shock is 
relativistic $\bar{\gamma}_3 \sim (n_1/n_4)^{1/4}(\eta/2)^{1/2} \gg 1$
and it considerably decelerates the shell material
$\gamma_3 \sim (n_4/n_1)^{1/4}(\eta/2)^{1/2} \ll \eta$.
Once $\gamma_3$ is determined, the density and the pressure in the
shocked shell region are given by $n_3 \sim (4\bar{\gamma}_3+3)n_4$ and
$p_3 \sim 4\gamma_3^2n_1m_pc^2/3$. 

In both cases, the time it takes for the reverse shock to cross a
distance $dx$ in the shell material can be given in a similar form 
up to a constant factor (Kobayashi, Piran \& Sari 1999),   
\begin{equation}
dR/c \propto \eta (n_4/n_1)^{1/2}~dx/c,
\label{eq:delta}
\end{equation}
where $R$ is the radius of the shell. Since the motion of the shell is
highly relativistic, we can regard $R/c$ as time in laboratory frame.  

When a shell is ejected from a source, it has a high density $n_4 \gg
\eta^2 n_1$, so the reverse shock is initially Newtonian. However,
as the shell expands, the shell density decreases. There is a
possibility that the shock becomes relativistic during it is crossing 
the shell. Sari and Piran (1995) showed that, using the Sedov length 
$l = (3E/4\pi n_1m_pc^2)^{1/3}$, if the shell is thick: $\Delta_0
>l/2\eta^{8/3}$, the reverse shock becomes relativistic at 
$R_N =l^{3/2}/\Delta_0^{1/2}\eta^2$ before the crossing at 
$R_\Delta = l^{3/4}\Delta_0^{1/4}$. If the shell is thin: $\Delta_0 <
l/2\eta^{8/3}$, it is dense enough to keep the reverse shock Newtonian.
The shock becomes mildly relativistic only when it just crosses the
shell at $R_\gamma=l/\eta^{2/3}$.  

The typical burst energy is about $10^{52}$
ergs. The ISM density has a typical values of 
$5$~protons/cm$^3$. The thick solid line in figure 
\ref{fig:eta_delta} separates the thick shell case (upper right) and the
thin shell case (lower left). According to the internal shocks model
the duration of a GRB $T$ is given by the shell width $\Delta_0/c$, the
thick shell cases correspond to relatively long bursts. 

\subsection{the Thick Shell Case}  

In a thick shell case the reverse shock becomes relativistic at $R_N$
and it begins to decelerate the shell material. Since the shell density 
decreases with radius as  $n_4 \sim n_1 l^3/\eta^2\Delta_0R^2$,  using
equation \ref{eq:delta}, one finds that the number of shocked electrons
$N_e$ is proportional to $R^2$. The scalings of the hydrodynamic
variables in terms of the observer time $t=R/2c\gamma_3^2$ are 
\begin{eqnarray}
\gamma_3 &\sim& 
\paren{l/\Delta_0}^{3/8}\paren{4ct/\Delta_0}^{-1/4}~;~
n_3 \sim 8 \gamma_3^3 n_1/\eta \propto t^{-3/4},\\
p_3 &\sim& 4\gamma_3^2 n_1m_pc^2/3 \propto t^{-1/2}~;~
N_e \sim N_0 ~ct/\Delta_0
\end{eqnarray}
where $N_0=E/\eta m_pc^2$ is the total number of electrons in 
the shell. Note that the Lorentz factor $\gamma_3$ does not
depend on its initial value $\eta$ after the reverse shock becomes
relativistic. 

After the shock crosses the shell at $t= \Delta_0/c$, the profile of the
forward shocked 
ISM begins to approach the Blandford-McKee (BM) solution (Kobayashi,
Piran \& Sari 1999). Since the shocked shell is located not too far
behind the forward shock, it roughly fits the BM solution. The
author and Sari (1999) numerically showed that the evolution is well
approximated by the BM solution if the relativistic reverse shock can
heat the shell to a relativistic temperature. Then, we apply the BM
scalings to the evolution of the shocked shell. The number of the shocked
electrons is constant after the shock crossing because no electron is
newly shocked.     
\begin{equation}
\gamma_3 \propto t^{-7/16}, ~n_3\propto t^{-13/16}, 
~p_3\propto t^{-13/12},
~N=\mbox{constant}.
\label{eq:BMscalings}
\end{equation}

\subsection{the Thin Shell Case}

In a thin shell case the reverse shock is too weak to decelerate the
shell effectively, the Lorentz factor of the shocked shell material
is almost constant during the shock propagates in the shell. Due to a
slight difference of the velocity inside the shell, the shell begins to
spread as $\Delta \sim R/\eta^2$ around $R = \eta^2\Delta_0$. Then,
the density ratio decreases as $n_4/n_1\sim (l/R)^3$.  The scalings
before the shock crosses the whole shell at $t_\gamma =l/2c\eta^{8/3}$
are given by    
\begin{equation}
\gamma_3 \sim \eta, 
    ~n_3 \sim 7n_1\eta^2(t/t_\gamma)^{-3},
    ~p_3 \sim 4\eta^2 n_1m_pc^2/3,
    ~N_e \sim N_0 (t/t_\gamma)^{3/2}.
\end{equation}
In the above thick shell case the spreading effect was not important
since it happens after the shock crossing.

The Newtonian reverse shock can not heat the shell to a relativistic
temperature which the BM solution assumes, then, we are not able to use
the BM solution. However, we derived scaling laws for a cold shocked
shell, assuming that the Lorentz factor is described by a power law
$\propto R^{-g}$ and that the shell expands adiabatically $p_3\propto
n_3^{4/3}$ with a sound speed $\sim \sqrt{p_3/n_3}$ in the shell's
comoving frame. We numerically showed that the scalings with $g\sim2$
fit the evolution (Kobayashi \& Sari 2000).  
\begin{equation}
\gamma_3 \propto t^{-2/5}, 
~n_3 \propto t^{-6/7}, ~p_3 \propto t^{-8/7},
~N=\mbox{constant}.
\end{equation}

\section{Light Curves of the Reverse Shock Emission}

We consider now the synchrotron emission from a reverse shocked 
shell. The shock accelerates electrons in the shell material into a
power law distribution: $N(\gamma_e)d\gamma_e\propto
\gamma_e^{-\hat{p}}d\gamma_e~(\gamma_e\ge\gamma_m)$. Assuming that a 
constant fraction $\epsilon_e$ and $\epsilon_B$ of the internal energy
go into the electrons and the magnetic field respectively, one finds 
that the typical random Lorentz factor of the electrons and the magnetic 
field evolve as $\gamma_m \propto \epsilon_e p_3/n_3$ and $B^2 \propto
\epsilon_Bp_3$. 

The spectrum is given by the broken power laws discussed in Sari, Piran
\& Narayan (1998). In this paper we neglect the self absorption since it
does not affect the optical radiation in which we are interested. Then,
it has two breaks at the typical synchrotron frequency $\nu_m \propto
B\gamma_3\gamma_m^2$ and at the cooling frequency $\nu_c \propto
1/B^3\gamma_3t^2$ which is  the synchrotron frequency of electrons that
cool on the dynamical time of the system. The peak flux is obtained at
the lower of the two frequencies. Let $N_e$ and $D$ be the total number
of the shocked electrons and the distance to the observer
respectively. The observed peak flux density evolves as  $F_{\nu,max}
\propto N_eB\gamma_3/D^2$. 

The spectra do not depend on the hydrodynamics of the shell. However,
the light curve at a fixed frequency depends on the temporal evolution
of the break frequencies $\nu_m$ and $\nu_c$ and the peak power
$F_{\nu,max}$. These depend on how $\gamma_3, n_3, p_3$ and $N_e$ scale
as a function of $t$. We will apply the adiabatic evolution discussed in
section 2 to the shell evolution. It is justified if the fraction of the
energy going to the electron is small $\epsilon_e \ll 1$ or if we are in
the regime of slow cooling $\nu_m < \nu_c$ where the electrons
forming the bulk of the population do not cool.    

\subsection{the Thick Shell Case} 
A reverse shock crosses a thick shell at $\sim \Delta_0/c$,
the peak time of the emission from the reverse shock is comparable to
the GRB duration $T$. Using the estimates in section 2 we 
obtain the break frequencies and the peak flux at the shock 
crossing time, 
\begin{eqnarray}
\nu_m&\sim& 7.3\times10^{14} 
~\epsilon_{e0}^2 
 \epsilon_{B0}^{1/2} 
 n_{1,5}^{1/2}
 \eta_{300}^2 ~\mbox{Hz}, \\ 
 \nu_c &\sim& 9.4\times10^{15} 
~\epsilon_{B0}^{-3/2}  E_{52}^{-1/2} n_{1,5}^{-1} 
 T_{100}^{-1/2}~\mbox{Hz}, \\
F_{\nu,max}&\sim& 0.3
~D_{28}^{-2} 
 \epsilon_{B0}^{1/2} 
 E_{52}^{5/4} n_{1,5}^{1/4}
\eta_{300}^{-1}
T_{100}^{-3/4}~\mbox{Jy},
\end{eqnarray}
where we have scaled the parameters as $E_{52}=E/10^{52}$ergs,
$\epsilon_{e0}=\epsilon_e/0.6$, $\epsilon_{B0}=\epsilon_B/0.01$,
$n_{1,5}=n_1/5~\mbox{cm}^{-3}$, $\eta_{300}=\eta/300$,
$T_{100}=T/100$ sec and $D_{28}=D/10^{28}$cm. The
equipartition values $\epsilon_e=0.6$, $\epsilon_B=0.01$ and
$n_1=5$~protons/cm$^3$ are inferred for GRB~970508 (Wijers \& Galama,
1999; Granot, Piran \& Sari 1999).  The scalings before and after 
the shock crossing are given by 
\begin{eqnarray}
t < T &:& ~\nu_m = \mbox{constant}, ~\nu_c \propto t^{-1}, 
~F_{\nu,max} \propto t^{1/2},\\
t > T &:& ~\nu_m \propto t^{-73/48}, ~\nu_c \propto t^{1/16}, 
~F_{\nu,max} \propto t^{-47/48}.
\end{eqnarray}
It is interesting that $\nu_m$ is constant during the shock crossing.  

\begin{figure}[b!] 
 \centerline{\epsfig{file=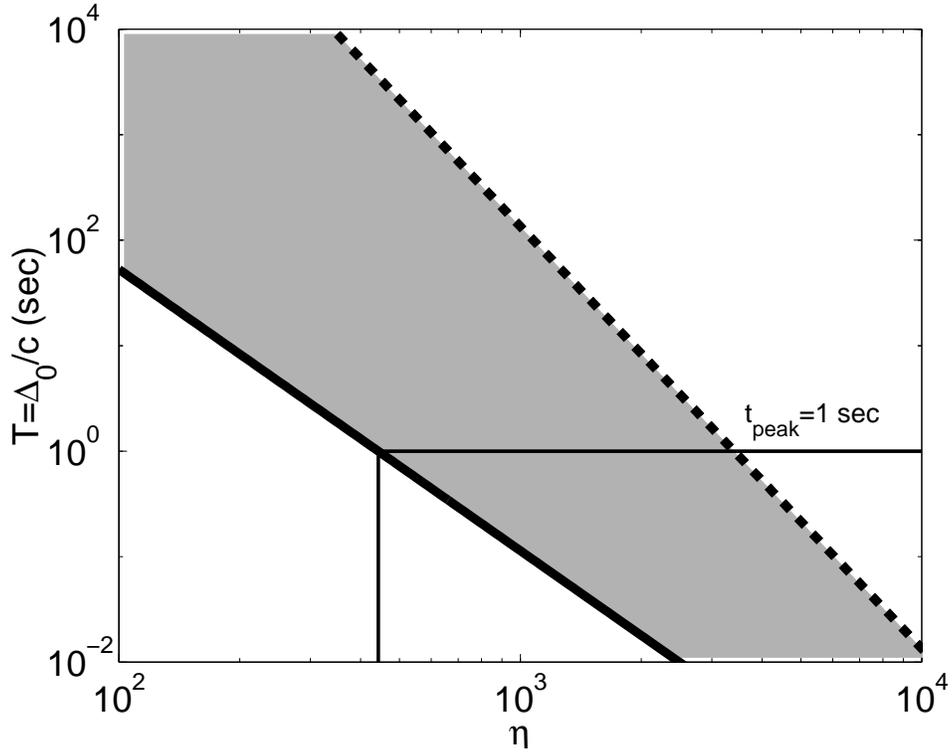,width=5in}} \vspace{10pt}
 \caption{Allowed parameter region: 
The thick solid line separates the thick shell case (upper right) and 
the thin shell case (lower left). The slow cooling region for the thick
shell case is lower left of the dashed line. That for the thin shell is
$\eta<8\times10^{4}$.
The peak time of the reverse shock emission is 
$t_{peak}=\mbox{max}[t_\gamma, T]$, the thin solid line depicts
$t_{peak}=1$~sec. 
$E=10^{52}$ergs, $n_1=5$ protons/cm$^3$, $\epsilon_e=0.6$ and 
$\epsilon_b=0.01$ are assumed.
} 
 \label{fig:eta_delta}
 \end{figure}

The spectrum is the slow cooling throughout the
evolution if $\nu_m < \nu_c$ at $T$. The shaded region in figure
\ref{fig:eta_delta} shows the corresponding parameter region.
The flux at a given frequency $\nu$ evolves as      
\begin{eqnarray}
F_{\nu}(t<T) &\propto& \left\{
                \begin{array}{@{\,}ll}
t^{1/2}       & \nu < \nu_c  \\   
\mbox{constant} & \nu > \nu_c 
                \end{array}
                \right. 
\label{fig:slowbefore}\\
F_{\nu}(t>T) &\propto& \left\{
                \begin{array}{@{\,}ll}
  t^{-17/36}             & \nu < \nu_{m}  \\          
  t^{-(73\hat{p}+21)/96} & \nu_m< \nu < \nu_{cut}  \\
  0                      & \nu > \nu_{cut}
                \end{array}
                \right. 
\end{eqnarray}
The flux at a frequency above $\nu_c$ disappears at $T$ because
no electron is shocked anymore. This cut off frequency $\nu_{cut}$
decreases as $\nu_c(T)(t/T)^{-73/48}$ due to the adiabatic
expansion of the fluid. The index  $-(73\hat{p}+21)/96$ is $\sim -2.1$
for the standard choice $\hat{p}=2.5$.

\begin{figure}[b!] 
 \centerline{\epsfig{file=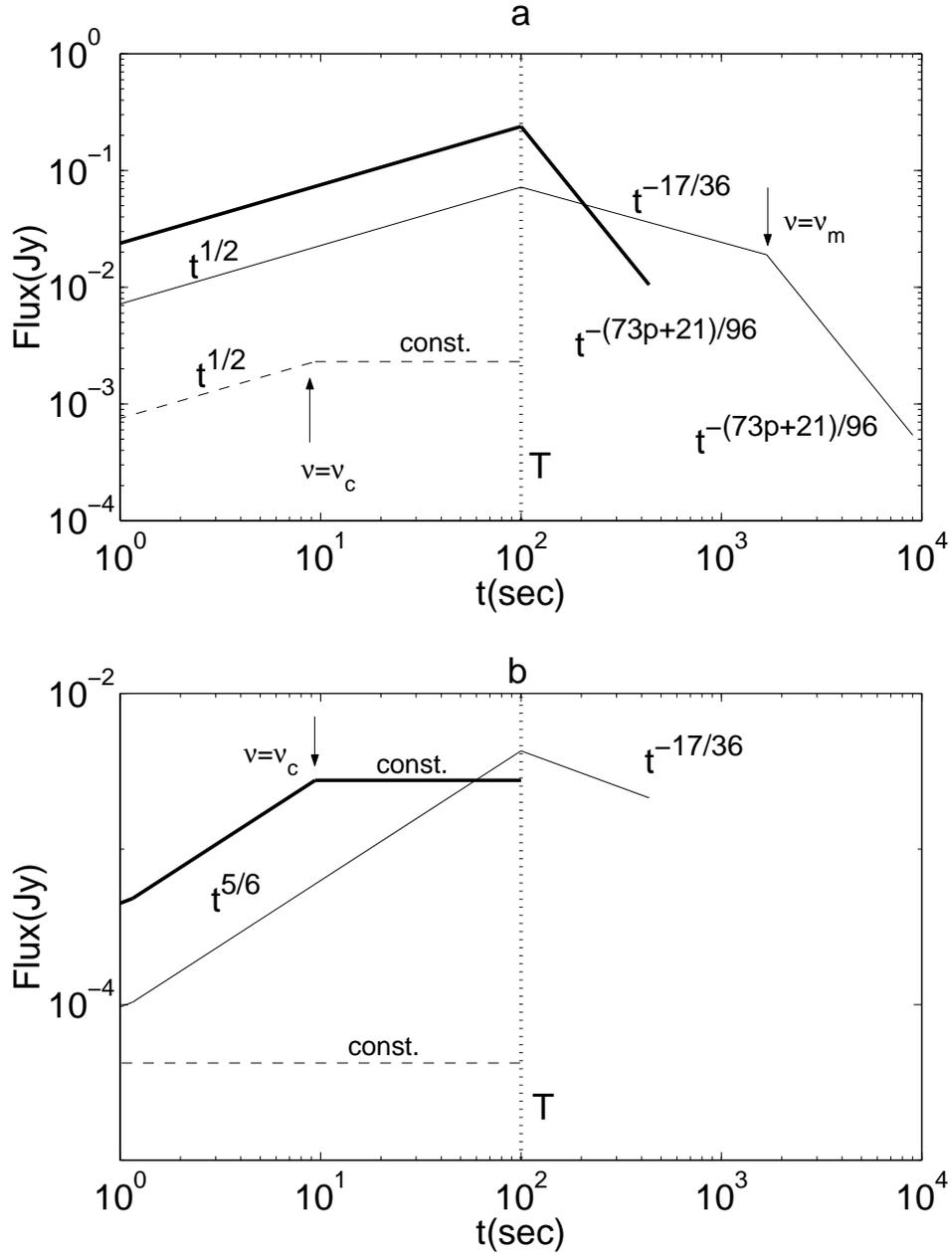,width=5in}} \vspace{10pt}
 \caption{
Light curves: thick shell case.
(a) Slow cooling case:  $\eta=300$.
thin solid ($\nu=10^{13}$Hz$<\nu_m(T)$),
thick solid ($\nu_m(T)<\nu=10^{15}$Hz$<\nu_c(T)$)
and dashed ($\nu=10^{17}$Hz$>\nu_c(T)$).
(b) Fast cooling case : $\eta=10^4$.
thin solid ($\nu=10^{15}$Hz$<\nu_c(T)$),
thick solid($\nu_c(T)<\nu=10^{17}$Hz$<\nu_m(T)$) 
and dashed ($\nu=10^{19}$Hz$>\nu_m(T)$).
$\nu=\nu_m$ and $\nu=\nu_c$ show the times when
the observe frequency $\nu$ is crossed by $\nu_m$ and $\nu_c$ 
respectively. 
$E=10^{52}$ergs, $n_1=5$ protons/cm$^3$, $\epsilon_e=0.6$, 
$\epsilon_b=0.01$ and $T=100$ sec are assumed.
} 
 \label{fig:lc_long}
 \end{figure}

The light curves are different among three frequency regimes separated
by the two frequency, $\nu_m$ and $\nu_c$ at $T$. The
typical light curves are shown in figure \ref{fig:lc_long}a. The flux
initially increases at all frequencies as $t^{1/2}$, but the high
frequency light curve flats when the cooling frequency crosses the given
frequency. At $T$, the flux begins 
to decay as $t^{-17/36}$ or $t^{-(73\hat{p}+21)/96}$. Finally it
vanishes when $\nu_{cut}$ crosses the given frequency $\nu$ at 
\begin{equation}
t\sim 700
~\epsilon_{B0}^{-72/73} 
 E_{52}^{-24/73} n_{1,5}^{-48/73}
 T_{100}^{49/73}
 \paren{\frac{\nu}{5\times10^{14}~\mbox{Hz}}}^{-48/73}~\mbox{sec}. 
\end{equation}

With a parameter set in the upper right region of the dashed line in figure
\ref{fig:eta_delta}, the spectrum changes from the slow cooling to the
fast cooling during the shock crossing.
The light curves are different among three frequency regimes separated
by $\nu_m$ and $\nu_c$ at $T$. The typical light curves are shown
in figure \ref{fig:lc_long}b. We neglected the initial slow cooling
phase in the figure since the transition happens at very early time. The
flux at a given frequency above $\nu_c$ is constant and below  $\nu_c$
it increases as $t^{5/6}$. When the shock crosses the shell, above
$\nu_c$ it vanishes. After that, below $\nu_c$ it decreases as
$t^{-17/36}$ until $\nu_{cut}$ crosses the given frequency. 

\subsection{the Thin Shell Case}

As a thin shell case, the shell width $\Delta_0/c$ should be smaller
than $t_\gamma$. Then, separation is expected between the GRB and the 
peak of the emission. The corresponding parameter region is lower left
of the solid line in figure \ref{fig:eta_delta}. The break frequencies
and the peak flux at $t_\gamma \sim3~E_{52}^{1/3}n_{1,5}^{-1/3} 
\eta_{300}^{-8/3}~\mbox{sec}$ are given by   
\begin{eqnarray}
\nu_m(t_\gamma) &\sim& 9.6\times10^{14}
~\epsilon_{e0}^2 
 \epsilon_{B0}^{1/2} n_{1,5}^{1/2}
 \eta_{300}^2 ~\mbox{Hz}, \\ 
\nu_c(t_\gamma) &\sim& 4.0\times10^{16}
~\epsilon_{B0}^{-3/2}  E_{52}^{-2/3}n_{1,5}^{-5/6} 
 \eta_{300}^{4/3}~\mbox{Hz}, \\
F_{\nu,max}(t_\gamma) &\sim&5.2 
~D_{28}^{-2}
 \epsilon_{B0}^{1/2} 
 E_{52} n_{1,5}^{1/2} 
 \eta_{300}~\mbox{Jy}.
\end{eqnarray}
Note that these do not depend on the initial shell width
$\Delta_0$ itself though $\Delta_0/c$ should be smaller than $t_\gamma$. 
The scalings before and after $t_\gamma$ are as follows, 
\begin{eqnarray}
t < t_\gamma&:& ~\nu_m \propto t^6, ~\nu_c \propto t^{-2}, 
~F_{\nu,max} \propto t^{3/2},\\
t > t_\gamma&:& ~\nu_m \propto t^{-54/35}, ~\nu_c \propto t^{4/35}, 
~F_{\nu,max} \propto t^{-34/35}.
\end{eqnarray}

\begin{figure}[b!] 
 \centerline{\epsfig{file=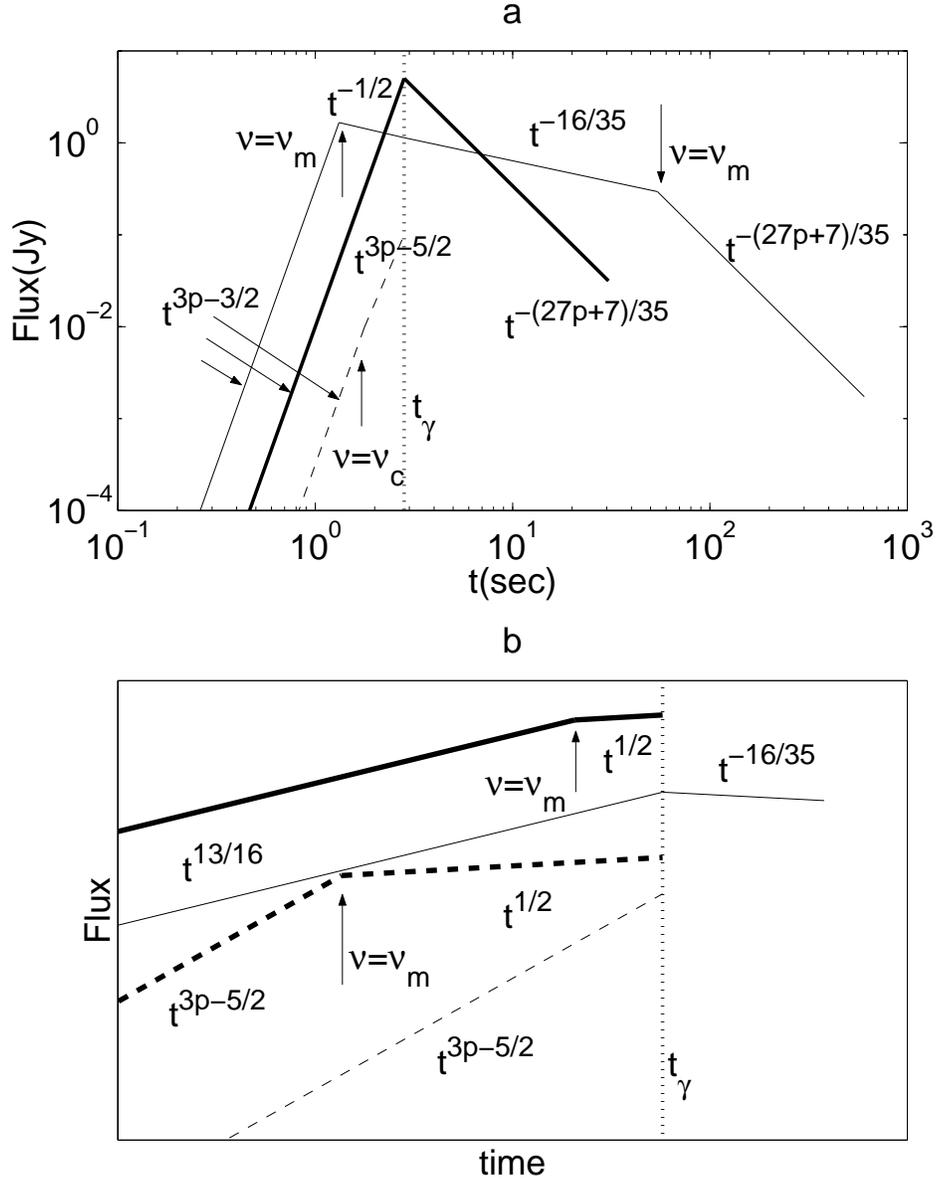,width=5in}} \vspace{10pt}
 \caption{
Light curve: thin shell case.
(a) Slow cooling case: $\eta=300$. 
Thin solid($\nu=10^{13}~\mbox{Hz}<\nu_m(t_\gamma))$, 
thick solid($\nu_m(t_\gamma)<\nu=10^{15}~\mbox{Hz}<\nu_c(t_\gamma)$) 
and dashed ($\nu=10^{17}~\mbox{Hz}>\nu_c(t_\gamma)$).
(b) Fast cooling case.
thin solid ($\nu<\nu_c(t_\gamma)$) , 
thin solid ($\nu_c(t_\gamma)<\nu< \nu_0$), 
thick dashed ($\nu_0 < \nu < \nu_m(t_\gamma)$) and 
thin dashed ($\nu > \nu_m(t_\gamma)$).   
$\nu=\nu_m$ and $\nu=\nu_c$ show the time when the observed frequency is
crossed by $\nu_m$ and $\nu_c$ respectively. 
$E=10^{52}$ergs, $n_1=5$ protons/cm$^3$, $\epsilon_e=0.6$ and 
$\epsilon_b=0.01$ are assumed.
} 
 \label{fig:lc_short}
 \end{figure}

If the spectrum is the slow cooling throughout the evolution, the
behavior of the light curves are different among the three frequency
regimes separated by $\nu_m$ and $\nu_c$ at $t_\gamma$.    
Figure \ref{fig:lc_short}a shows typical light curves.
The flux initially increases in all regimes very rapidly as 
$t^{3\hat{p}-3/2}(\sim t^6$ for $\hat{p}=2.5$). The slope changes when
$\nu_m$ or $\nu_c$ crosses the given frequency. After $t_\gamma$, the
flux decays as $t^{-16/35}$ or $t^{-(27\hat{p}+7)/35} ~(\sim t^{-2.1}$
for $\hat{p}=2.5$). The emission at a frequency $\nu < \nu_c(t_\gamma)$
disappears when the cut off frequency crosses it at 
\begin{equation}
t\sim 50
~\epsilon_{B0}^{-35/36} 
 E_{52}^{-8/81}n_{1,5}^{-283/324}
 \eta_{300}^{-146/81}
 \paren{\frac{\nu}{5\times10^{14}~\mbox{Hz}}}^{-35/54}~\mbox{sec}. 
\label{eq:dis}
\end{equation}

If $\nu_m$ is higher than $\nu_c$ at $t_\gamma$, 
the spectrum changes from the slow cooling to the fast cooling during
the shock crossing. However, it requires a large Lorentz factor
$\eta>8\times10^4~\epsilon_{e0}^{-2}\epsilon_{B0}^{-2}E_{52}^{-1}      
n_{1,5}^{-2}$. Then,  the transition happens only if the shell is
extremely thin (see figure \ref{fig:eta_delta}), then it is hard to 
detect such a prompt emission. 
We show the possible light curves for completeness. The behavior is
different among the following four frequency regimes: 
$\nu<\nu_c(t_\gamma)$,  
$\nu_c(t_\gamma)<\nu< \nu_0$,
$\nu_0 < \nu < \nu_m(t_\gamma)$ and 
$\nu > \nu_m(t_\gamma)$
where $\nu_0=\nu_m(t_0)=\nu_c(t_0)$ and $t_0$ is the transition time 
from the slow cooling to the fast cooling. The typical light curves 
are shown in figure \ref{fig:lc_short}b in which the very early slow
cooling phase is neglected. 

\section{Observations}
We in this section compare our estimates with the ROTSE observations.
First, we determine the parameters of the fireball of GRB~990123 
from the observations. Then, we make some comments on the lack of the
prompt optical emission from GRB~981121 and GRB~991223 in the context
of our model. 

\subsection{GRB~990123}
GRB~990123 is a very bright burst which fluence is about 100 times that 
of a median BATSE burst.  Absorption lines in the optical afterglow
gives a lower limit of the redshift $z>1.6$, the energy required to
produce the bright GRB is enormous, $\sim10^{54}$ ergs for an isotropic 
emission. ROTSE detected a strong optical flash during the burst
(Akerlof et al. 1999). About $50$ sec after the GRB trigger, it reached
to a peak of $\sim 1$ Jy and then decayed with a slope of a power law
index $\sim 2$. 

The fireball of GRB~990123 is likely to be a marginal case (Kobayashi \&
Sari 2000). However, such a marginal case behaves very much like the
thin shell case, since the shell heated by a mildly relativistic shock
becomes cold soon. The observed  decay is in good agreement with
the  theory $t^{-2.1}$. This means that the relation among $\nu_m$,
$\nu_c$ and the observed optical frequency $\nu_R \sim 5\times10^{14}$
Hz is $\nu_m < \nu_R <\nu_c$ at the peak time. Furthermore, the strong
optical emission implies that $\nu_R$ is close to $\nu_m$
(Sari \& Piran 1999b).  

It is well-known that the peak time $t_{peak}$ is sensitive to the
Lorentz factor of the shell, so we obtain $\gamma_3 \sim
180 ~E_{54}^{1/8}n_{1,5}^{-1/8}(t_{peak}/50~\mbox{sec})^{-3/8}$. 
Since the reverse shock is mildly relativistic, the Lorentz factor at 
the peak time should be close to the initial Lorentz factor which can be
estimated by using equation \ref{eq:dis}. The last detection of the
optical flash by ROTSE was $\sim 600$ sec after the GRB trigger. Using
the equality of the Lorentz factors, we obtains $\eta\sim270$ and
$n_1\sim0.2$ protons/cm$^3$. With these parameters the typical
synchrotron frequency at the peak time is $\sim 1.6\times10^{14}$ Hz
which is close to $\nu_R$ and it is consistent with the initial
assumption.      

Though we can not give a strong argument on the raising part of the 
light curve since the observation is sparse, the power law index
of the slope calculated from the first two ROTSE data is 3.4. The
sparseness can make only the index smaller, then the real index is at
least larger than that of a thick shell.

\subsection{GRB~981121 and GRB~981223}
\begin{figure}[b!] 
 \centerline{\epsfig{file=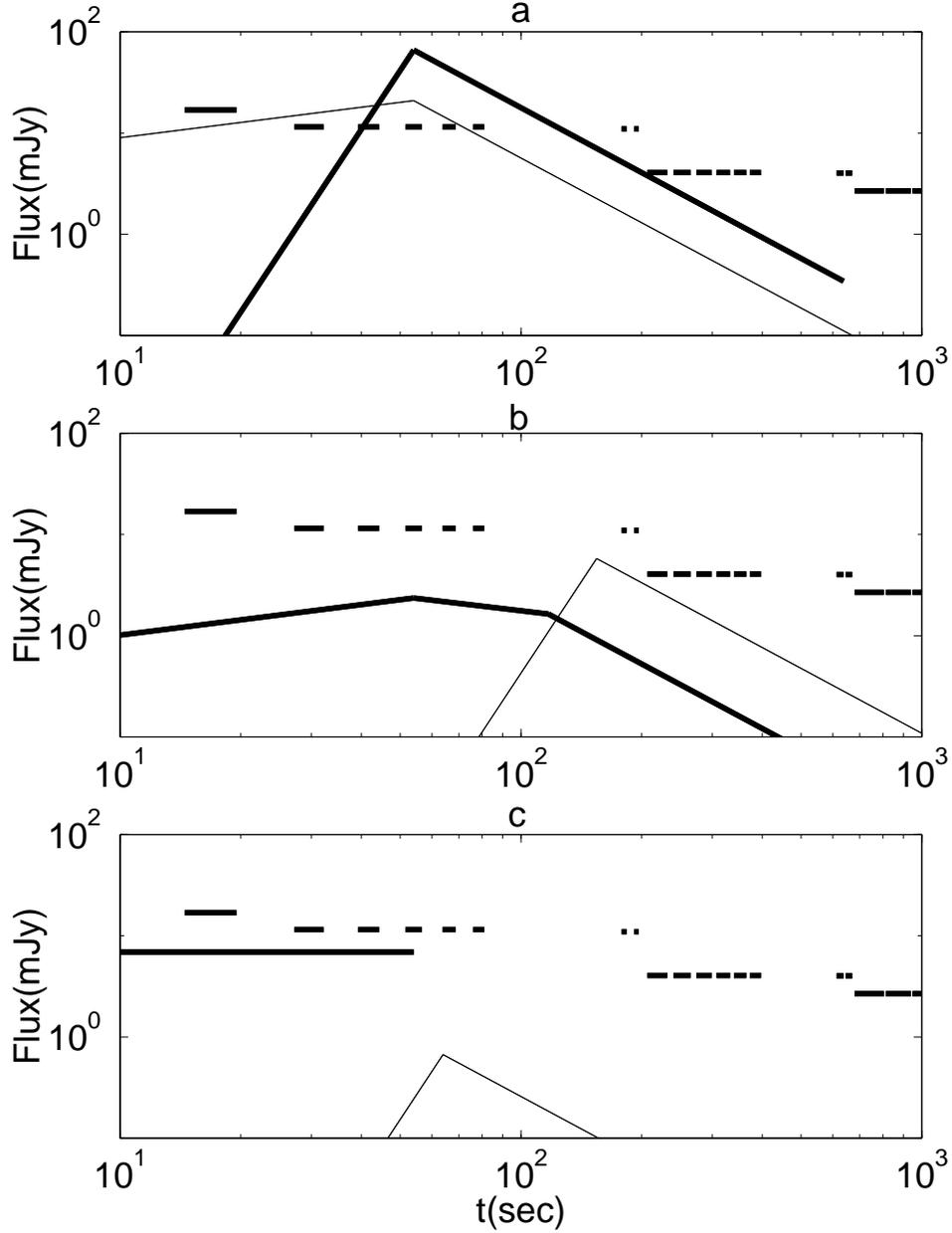,width=5in}} \vspace{10pt}
 \caption{ GRB~981121: the ROTSE detection thresholds (segments) and
 the theoretical light curves. 
(a) $E=10^{52}$ ergs (thin) and $E=10^{54}$ ergs (thick).
(b) $\eta=100$  (thin) and $\eta=1000$ (thick). 
(c) $n_1=10^{-3}$ protons/cm$^3$ (thin) and 
    $n_1=10^{6}$ protons/cm$^3$  (thick).
$n_1=0.2$ protons/cm$^3$,
  $\eta=270$, $E=10^{52}$ergs and $T=54$ sec are used if 
  the values are not specified.
The thresholds are calculated assuming that the ROTSE magnitude
  corresponds to that in the V band.  
} 
 \label{fig:981121}
 \end{figure}

\begin{figure}[b!] 
 \centerline{\epsfig{file=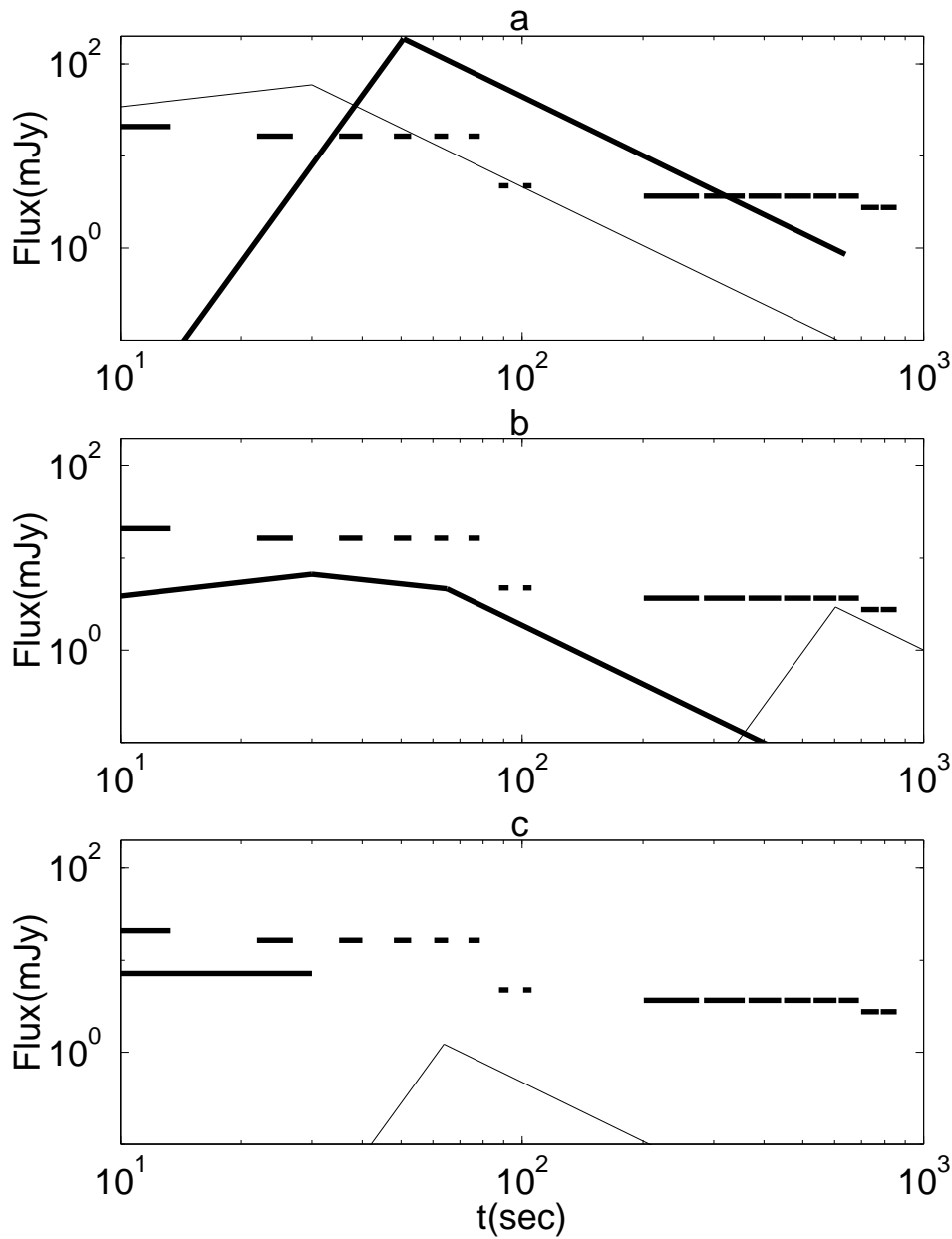,width=5in}} \vspace{10pt}
 \caption{GRB~981223: the ROTSE detection thresholds (segments) and
the theoretical light curves. 
(a) $E=10^{52}$ ergs (thin) and $E=10^{54}$ ergs (thick).
(b) $\eta=60$  (thin) and $\eta=1000$ (thick). 
(c) $n_1=10^{-3}$ protons/cm$^3$ (thin) and
    $n_1=10^{8}$ protons/cm$^3$ (thick).
$n_1=0.2$ protons/cm$^3$,
  $\eta=270$, $E=10^{52}$ergs and $T=30$ sec are used if 
  the values are not specified.
} 
 \label{fig:981223}
 \end{figure}

The theory succeeded for GRB~990123, but the optical flash was detected
only for it so far. Akerlof et al. (2000) reported no detections of the
optical flashes to six GRBs with localization errors of 1 deg$^2$ or
better. Especially, GRB~981121 and GRB~981223 are the most sensitive
bursts in the sample. If the optical flashes are correlated with the GRB 
fluences, the optical emission should be more than 2 mag over the 
ROTSE detection thresholds. The thick lines in figure \ref{fig:981121}a
and \ref{fig:981223}a show the light curves expected from the GRB
fluences,
$f_{GRB981121}=7.0\times10^{-6}$ ergs/cm$^{2}$, 
$f_{GRB981223}=1.3\times10^{-5}$ ergs/cm$^{2}$ and
$f_{GRB990123}=1.0\times10^{-4}$ ergs/cm$^{2}$
(Akerlof et al. 2000).

GRBs are produced by the internal shocks, while the afterglows and the
optical flashes are due to the external shocks. The lack of a direct
scaling between the GRB and the afterglow is an evidence of the internal
shocks model. However, the energy radiated in the GRB phase $E_{\gamma}$
is still correlated with the blast wave energy $E$ estimated from the
x-ray afterglow (Freedman and Waxman 1999). According to their analysis,
the ratios $E_{\gamma}/E$ of 13 events are in between 0.1 and 6, while
$E_\gamma$ ranges from $3\times10^{51}$ ergs to $10^{54}$
ergs. Furthermore, $E_{\gamma}/E\sim3$ of GRB~990123 is a relatively 
large value in the sample. Therefore, the scaling with the GRB fluence 
does not overestimate $D^{-2}E$ in general.  

GRB~990123 was an exceptionally energetic burst. The energies of
GRB~981121 and GRB~981223 should be considerably lower to explain the
lower fluences, otherwise the sources are extremely far, z=10 for
GRB~981121 and z=6.5 for GRB~981223 assuming $h=0.65, \Omega_0=1$ and
$\lambda_0=0$. Hereafter we assume $E=10^{52}$ ergs for the two
bursts. Since GRB~990123 is a marginal case, a burst with a lower E and
a comparable $n_1$, $\eta$ and $T$ is classified into the thick shell
case in which $F_{\nu,max}$ is proportional to $E^{5/4}$ instead of
$E$. Then, the peak flux for $E=10^{52}$ ergs is lower by a factor of
$\sim 3$ than that just scaled by the GRB fluence. The thin lines in 
figure \ref{fig:981121}a and \ref{fig:981223}a depict the corrected
light curves which are still above the thresholds. The durations of
GRB~981121 and GRB~981223 are 54 sec and 30 sec respectively and
comparable to that of GRB~990123.

A possible solution to the problem is to assume that the reverse shock
energy is radiated at a non-optical frequency, $\nu_m \ll \nu_R$ or 
$\nu_m \gg \nu_R$. The typical frequency $\nu_m$ is proportional to
$\epsilon_e^2\epsilon_B^{1/2}n_1^{1/2}\eta^2$, but the values of
$\epsilon_e$ and $\epsilon_B$ are determined by the micro-physics and
are likely to be universal. The difference of $\nu_m$ should be due to
that $n_1$ and $\eta$ of the bursts are different from the
``canonical'' values, $n_1=0.2$ protons/cm$^3$ and $\eta=270$.

\begin{figure}[b!] 
 \centerline{\epsfig{file=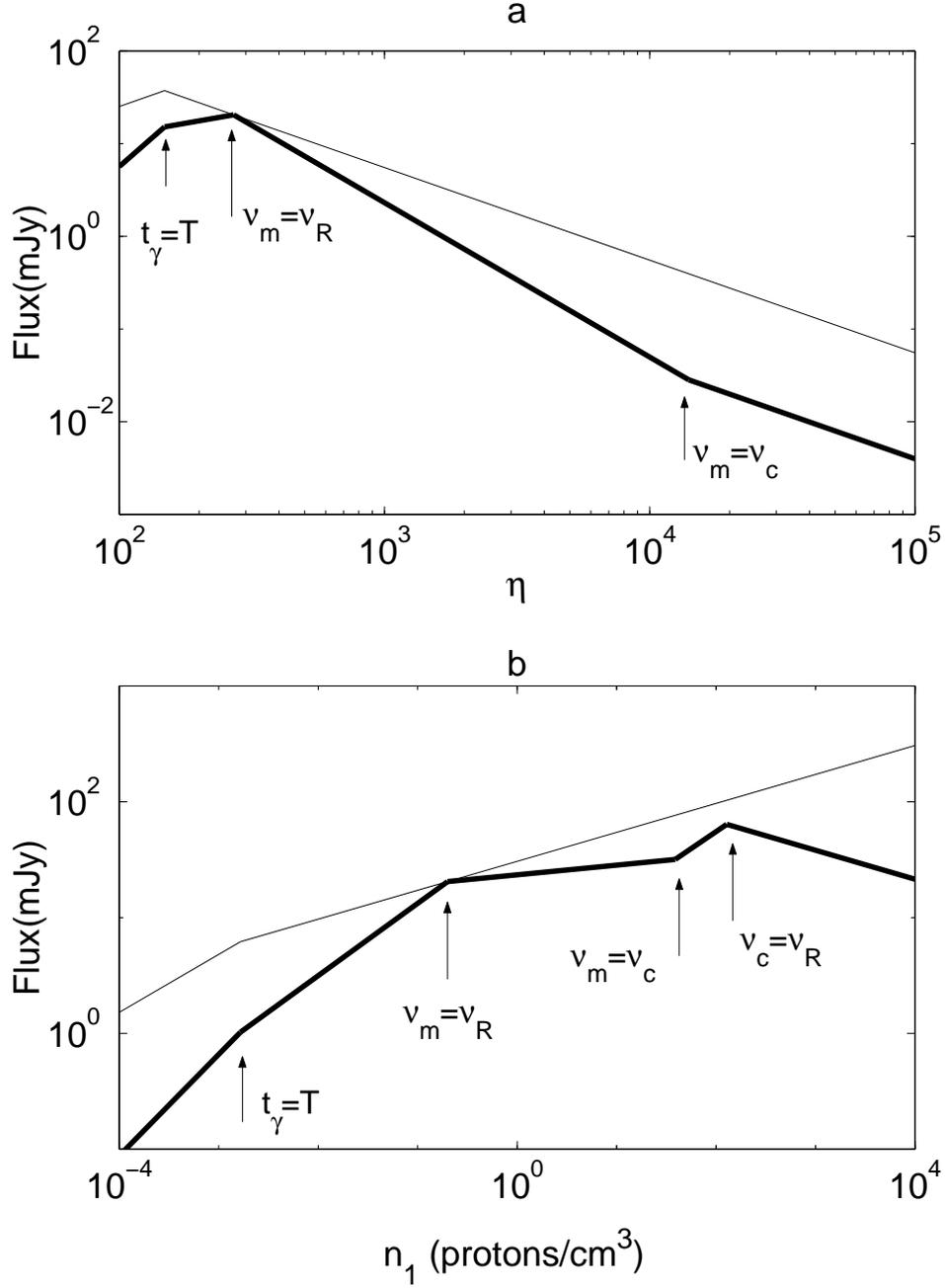,width=5in}} \vspace{10pt}
 \caption{ GRB~981121: the optical flux $F_{\nu_R}$ 
(thick) and the peak power $F_{\nu,max}$ (thin) at the peak time
max[$t_\gamma, T$].
$\nu_m=\nu_R$ shows the points for the canonical parameters.
} 
 \label{fig:peak}
 \end{figure}

If $n_1$ and $\eta$ are smaller than the canonical values, $\nu_m$ is
lower than $\nu_R$ since $\nu_m\sim\nu_R$ for GRB~990123. Using the
normalization by the GRB fluence, the peak flux is 
\begin{equation}
F_{\nu_m<\nu_R<\nu_c}(T) \sim 1.0 ~f_{0} E_0^{1/4}
T_{0}^{-3/4} n_0^{\hat{p}/4}\eta_0^{\hat{p}-2} ~\mbox{Jy}
\end{equation}
where the subscript 0 denotes that the parameters are scaled by the
values of GRB~990123, $f_0=f/10^{-4}$ ergs cm$^{-2}$, $E_0=E/10^{54}$ 
ergs, $T_{0}=T/50$ sec, $n_0=n_1/0.2$ protons cm$^{-3}$ and
$\eta_0=\eta/270$.  The fireball is classified into the thin shell case
if we assume very small $n_1$ or $\eta$, thus the dependence of the peak
flux on $n_1$ and $\eta$ changes to $F_{\nu_R}(t_\gamma)\propto
n_0^{(\hat{p}+1)/4}\eta_0^{\hat{p}}$.  $F_{\nu_R}$ and $F_{\nu,max}$ at
the peak time are plotted in figure \ref{fig:peak} as functions
of $\eta$ and $n_1$ in the case of GRB~981121. The no detections by ROTSE
give upper limits on $n_1$ and $\eta$. Assuming $n_1=0.2$ protons/cm$^3$,
we get $\eta<135$ for GRB~981121 and $\eta<65$ for GRB~981223, or assuming
$\eta=270$, $n_1<0.07$ protons/cm$^3$ for GRB~981121 and $n_1<0.02$
protons/cm$^3$ for GRB~981223. The light curves with a low $n_1$ or 
$\eta$ are shown as the thin lines in figure \ref{fig:981121}b,c and
\ref{fig:981223} b,c.

If we assume a larger $n_1$ or $\eta$ than the canonical values, the
peak flux is given by 
\begin{equation}
F_{\nu_R<\nu_m}(T) \sim 1.0 ~f_0 E_0^{1/4}T_0^{-3/4}
n_0^{1/12}\eta_0^{-5/3} ~\mbox{Jy}.
\end{equation}
The spectrum changes to the fast cooling if $n_1$ or $\eta$ is very
large, the dependence on $n_1$ and $\eta$ changes to 
$F_{\nu_R} \propto n_0^{7/12}\eta_0^{-1}$. For a further
larger $n_1$, $\nu_c$ can be lower than $\nu_R$, then the dependence 
becomes $n_0^{-1/4}$.  The no detections give lower limits, 
assuming $n_1=0.2$ protons/cm$^3$, $\eta>400$ for GRB~981121 and 
$\eta>600$ for GRB~981223, or assuming $\eta=270$,
$n_1=2\times10^5$ protons/cm$^3$ for GRB~981121 and 
$n_1=4\times10^6$ protons/cm$^3$ for GRB~981223. The light curves for a
large $n_1$ or $\eta$ are shown as the thick lines in figure
\ref{fig:981121}b,c and \ref{fig:981223}b,c. 

A large ISM density is an unlikely reason to explain the no detections,
because we need to require several order larger density than the
canonical and with the large density the peak power $F_{\nu,max}$ itself
is very large (see figure \ref{fig:peak}b). 
However, there is a possibility of the extinction. Though we have
normalized the optical flux according to the GRB fluence, gamma-rays do
not suffer any kind of extinction. A half of x-ray afterglow bursts do 
not have optical afterglows, it might be due to the extinction. The
reverse shock radiates at a closer region to the inner engine, it might
be crucial.  Within the six bursts to which the ROTSE group reported the
no detections, only the location of GRB~980329 was determined precisely
by BeppoSAX, and the optical afterglows were observed hours
later. However, the ROTSE observation on this burst was not so 
sensitive. Future observations will provide some information to this
effect.  

\section{Conclusions}
We have constructed the full light curves of the reverse shock emission
for a short burst (thin shell) and a long burst (thick shell). The
typical synchrotron frequency increases rapidly as $t^6$ in the 
thin shell case while it is constant in the thick shell case.
For a plausible moderate Lorentz factor, upto a few thousand, the
synchrotron spectrum is the slow cooling throughout the evolution,
with which spectrum we find that the flux must rise initially steeply
as $t^{3\hat{p}-5/2}$ or $t^{3\hat{p}-3/2}$ in the thin shell case, and
slowly as $t^{1/2}$ in the thick shell case.   

The rise ends when the reverse shock crosses the shell. Only the
exception is the low frequency emission from a thin shell which already
begins to decrease when the rapidly changing typical frequency crosses
the observe one.  In the decay phase, the light curves are similar for
both cases, though the hydrodynamics are very different. Then, the
detection before the peak should be more useful to give a constraint on
the initial properties of the fireball.  

The prompt optical emission from GRB~990123 is well described by the
reverse shock emission. The observations enabled us to determine
the initial Lorentz factor and the ISM density. We found that 
GRB~990123 is a ``luck'' burst. Besides the exceptionally large
energy,  it has an optimized Lorentz factor to produce a bright 
optical flash. As the initial Lorentz factor increases, 
the peak power, $F_{\nu,max}\propto \gamma_3^2/\eta$, initially
rises. However, with a moderate initial Lorentz factor
the evolution changes to the thick shell case in which the 
Lorentz factor at the peak time  no longer depends on the
initial Lorentz factor. Then, the peak power drops since the number of
the electrons in the shell continues to decrease. Therefore, the
marginal case gives the  brightest emission. GRB~990123 is the marginal
case and the typical frequency $\nu_m\propto n_1^{1/2}\eta^2$ just
comes to the optical band.

The lack of the prompt optical detections by ROTSE for GRB~981121 and
GRB~981223 does not give strong constraints on the initial Lorentz 
factors or the ISM densities. If the Lorentz factor is slightly 
different from that of GRB~990123, the peak flux becomes lower than the
ROTSE thresholds. It is also possible to explain the no detections by a
lower ISM density. 

\vspace{0.5cm}
The author thanks Re'em Sari for helpful discussions, and Robert Kehoe  
for providing with the ROTSE data. This work was supported by the Japan
Society for the Promotion of Science.   
\vspace{1cm}
\newpage

\noindent {\bf References}\newline
Akerlof,C.W. et al. 1999, Nature, 398, 400.\newline
Akerlof,C.W. et al. 2000, ApJ, 532, L25.\newline
Blandford,R.D. \& McKee,C.F. 1976, Phys of Fluids, 19, 1130.\newline
Freedman, D.L. and Waxman, E. 1999, astro-ph/9912214.\newline
Granot, J., Piran, T. \& Sari, R., 1999, ApJ, 527, 236.\newline
Kobayashi,S., Piran,T. \& Sari,R. 1999, ApJ, 513, 669.\newline
Kobayashi,S. \& Sari,R. 2000, ApJ, 543, in press.\newline
M\'{e}szaros,P. \& Rees,M.J. 1997, ApJ, 476, 231.\newline   
Sari,R. \& Piran,T. 1995, ApJ, 455, L143.\newline
Sari,R. \& Piran,T. 1999a, ApJ, 520, 641.\newline 
Sari,R. \& Piran,T. 1999b, ApJ, 517, L109.\newline 
Sari,R. \& Piran,T. \& Narayan,R. 1998, ApJ, 497, L17.\newline
Wijers,R.A.M.J. \& Galama, T.J. 1999, ApJ, 523, 177.\newline
\newpage
\end{document}